\documentclass[letterpaper]{article}
\usepackage[preprint]{aaai2027}
\usepackage[hyphens]{url}
\usepackage{graphicx}
\usepackage{natbib}
\usepackage{caption}
\usepackage{booktabs}
\usepackage{amsmath,amssymb}
\usepackage{placeins}

\newcommand{\method}{GeneGeoFlow}
\newcommand{\methodsp}{GeneGeoFlow-Prop}

\title{Control-Anchored Residual Flow Matching Conditioned on Gene Geometry for Virtual Cell Perturbation Modeling}
\author{
Quanquan Li\textsuperscript{\rm 1},
Yihe Chi\textsuperscript{\rm 1},
Liuyang Song\textsuperscript{\rm 2},
Hongbo Zhang\textsuperscript{\rm 2},
Jingyu Li\textsuperscript{\rm 3},
Xidong Xi\textsuperscript{\rm 1},\\
Conghua Wei\textsuperscript{\rm 1},
Yijie Sun\textsuperscript{\rm 4},
Yu Chen\textsuperscript{\rm 1},
Xin Liu\textsuperscript{\rm 1},
Qi Hu\textsuperscript{\rm 1},
Jing Ke\textsuperscript{\rm 5},
Guitao Cao\textsuperscript{\rm 1}\corresponding
}
\affiliations{
\textsuperscript{\rm 1}East China Normal University\\
\textsuperscript{\rm 2}Peking University\\
\textsuperscript{\rm 3}University of Science and Technology of China\\
\textsuperscript{\rm 4}Nanjing University\\
\textsuperscript{\rm 5}Shanghai Jiao Tong University
}

\begin{document}
\maketitle

\begin{abstract}
A central task in virtual cell modeling is predicting single-cell transcriptional responses to unseen genetic perturbations and drug combinations, and biological networks provide valuable priors on gene relationships.
Existing graph-based models commonly use the same network to structure gene representations and mediate intergene interactions, thereby implicitly treating stable associations as perturbation-response pathways.
Gene Ontology and control-derived coexpression networks encode relatively stable relationships rather than intervention-specific response directions or magnitudes.
We therefore propose \method{}, which conditions a control-anchored residual flow on gene-wise geometry derived from biological networks to learn intervention-specific transcriptional responses.
\method{} derives multi-scale spectral coordinates from Gene Ontology and control-derived coexpression networks.
A perturbation-conditioned, gene-wise gating module selects relevant structural scales and network sources, yielding intervention-specific gene geometry.
The resulting geometry conditions a control-anchored residual flow without explicitly propagating target-derived signals along the graph.
Condition-wise optimal transport couples unpaired control and perturbed populations for training, while a Delta-correlation objective aligns the predicted and observed condition-level expression-shift directions.
\method{} achieves Pearson Delta scores of $0.8979$ on the Norman additive benchmark and $0.9088$ on five held-out drug combinations in the fixed ComboSciPlex test split.
These results support perturbation-conditioned gene geometry as an effective structural prior for intervention-specific response prediction, without conflating stable gene relationships with response propagation.
\end{abstract}

\section{Introduction}

\begin{figure*}[t!]
\centering
\includegraphics[width=0.95\textwidth]{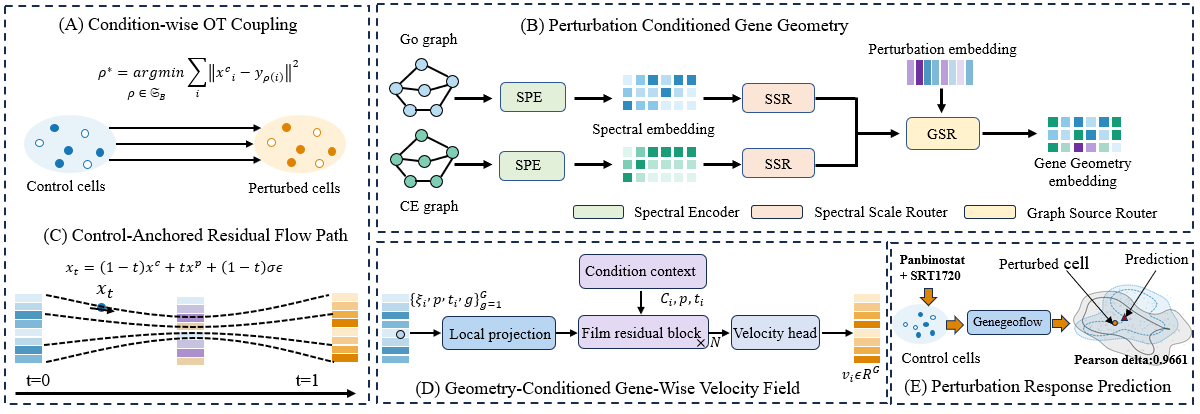}
\caption{Overview of \method{}. \textbf{(A)} Condition-wise optimal transport couples unpaired control and perturbed populations to construct training endpoints. \textbf{(B)} Multi-scale spectral coordinates from the GO and control-derived coexpression graphs are selected by the Spectral Scale Router and fused by the perturbation-conditioned Graph Source Router, yielding gene-wise geometry $Z_p$. \textbf{(C)} The control-anchored residual path interpolates between each coupled control--perturbed pair to form the intermediate state $x_t$. \textbf{(D)} A shared local projection maps $(x_t,x^c,s_p,Z_p)$ to gene-wise features; the condition context modulates three FiLM residual blocks, and the velocity head outputs one scalar velocity per gene. \textbf{(E)} Integrating the learned cell-wide velocity field predicts the perturbed population from control cells. Panobinostat+SRT1720 is shown only as a running illustration: its $0.9661$ Pearson Delta is condition specific, whereas aggregate claims are evaluated across all held-out conditions.}
\label{fig:method_overview}
\end{figure*}

The space spanned by genetic perturbations and drug combinations far exceeds what single-cell experiments can measure~\cite{norman2019exploring,lotfollahi2023cpa}.
Moreover, destructive single-cell sequencing cannot provide paired pre- and post-intervention observations of the same cell, leaving available data sparse across conditions and unpaired across cells~\cite{bunne2023cellot}.
Virtual-cell models must therefore learn extrapolatable intervention responses from limited population-level observations.
Consider the held-out Panobinostat+SRT1720 combination in Figure~\ref{fig:method_overview}: starting from a control-cell population, the model must use response patterns learned from other training conditions to predict the combination-specific response distribution and gene-expression shift.
Thus, the central problem is to learn from sparse, unpaired populations a response mapping that generalizes to unseen interventions.

To enable such cross-intervention extrapolation, existing virtual-cell models learn transferable perturbation representations~\cite{lotfollahi2019scgen,lotfollahi2023cpa}, transport unpaired cell populations~\cite{bunne2023cellot,driessen2026cmonge}, or model response distributions with flow, bridge, or diffusion dynamics~\cite{klein2025cellflow,chi2026departures,zhang2026scdfm,huang2026perturbdiff}.
Graph-based predictors such as GEARS further introduce biological priors, but use gene-relation graphs both to structure gene representations and to mediate intergene interactions~\cite{roohani2023gears}.
Yet GO annotations encode curated functional relations~\cite{ashburner2000geneontology}, whereas control-derived coexpression reflects statistical associations among unperturbed cells; neither specifies the direction or magnitude of transcriptional change under a particular intervention.
Using the same relations as response paths therefore conflates stable association with intervention-specific propagation.
Thus, the key question is not how to propagate perturbations over a graph, but how to use biological networks as structural context while learning response dynamics from data.

Answering this question requires what we term structure--dynamics separation: biological networks should provide gene-wise structural context, whereas intervention-specific response dynamics should be learned from data.
This separation presents three technical challenges.
First, a fixed graph source or spectral scale imposes the same structural view on every intervention, ignoring the condition-dependent roles of GO and control-derived coexpression.
Second, control and perturbed cells have no one-to-one correspondence; arbitrary pairing confounds intervention effects with differences in baseline cell state~\cite{bunne2023cellot,chi2026departures}.
Finally, cell-wise learning objectives can fit plausible individual cells without preserving the direction of condition-level gene-expression changes~\cite{ahlmann2025baselines,vinas2025systema}.
The model must therefore jointly support intervention-adaptive structure selection, training couplings between unpaired populations, and condition-level response alignment.

Guided by this principle, we introduce \method{}, which operationalizes structure--dynamics separation by using biological networks as gene-wise structural coordinates and an independent response field to learn intervention-specific transcriptional changes.
The model first extracts multi-scale spectral coordinates from GO and control-derived coexpression graphs.
Perturbation-conditioned Spectral Scale Router and Graph Source Router then select the relevant spectral scales and graph sources for each gene, yielding gene geometry.
Next, condition-wise optimal transport constructs training couplings between unpaired control and perturbed populations, and a control-anchored gene-wise residual flow learns intervention-specific population transitions conditioned on gene geometry.
Finally, a Delta-correlation objective aligns the directions of predicted and observed condition-level expression changes.

We evaluate \method{} primarily on the Norman strict genetic-perturbation holdout protocol and use the fixed ComboSciPlex drug-combination split for cross-domain validation.
On Norman strict holdout, gene geometry improves the five-fold mean Pearson Delta from $0.1450$ to $0.8153$ and outperforms the graph-free model in every fold; perturbation-conditioned geometry further improves over static geometry in matched folds, whereas explicit graph propagation yields no stable gain.
On the established Norman additive benchmark, \method{} achieves $0.8979$ Pearson Delta.
On five held-out ComboSciPlex drug combinations, \method{} achieves $0.9088$ Pearson Delta, compared with $0.8636$ for the explicit-propagation variant and $0.5169$ for the graph-free model.
Within these matched comparisons, the dominant improvement arises from perturbation-conditioned gene geometry, while the evaluated propagation operator provides no consistent additional gain.

Our contributions are summarized as follows:
\begin{enumerate}
  \item We identify a fundamental role entanglement in existing graph-based perturbation models: biological networks are used both to represent gene organization and to mediate intervention-response propagation. We introduce a structure--dynamics separation perspective, in which biological networks define gene-wise structural context while intervention-specific response dynamics are learned from data.
  \item We propose \method{}, which constructs perturbation-conditioned gene geometry through multi-scale routing over Gene Ontology and control-derived coexpression graphs and uses this geometry to condition a control-anchored, gene-wise residual flow. Condition-wise optimal transport and a Delta-correlation objective address unpaired cell populations and condition-level response alignment, respectively.
  \item Through matched comparisons with graph-free, static-geometry, and explicit-propagation controls on Norman and ComboSciPlex, we demonstrate that perturbation-conditioned gene geometry provides substantial improvements in predicting responses to unseen genetic perturbations and drug combinations.
\end{enumerate}

\section{Related Work}

\paragraph{Perturbation-conditioned gene geometry.}
Graph-based perturbation modeling has evolved from explicit relation propagation to module-aware and graph-conditioned response prediction.
GEARS propagates perturbation information over gene relations~\cite{roohani2023gears}, while scBIG organizes gene responses through module-inductive representations~\cite{ruan2026scbig}.
scDFM further incorporates a graph-aware backbone into distributional flow matching~\cite{zhang2026scdfm}.
In parallel, scGen, CPA, chemCPA, biolord, AdaPert, and PerturbedVAE encode interventions through transferable, compositional, disentangled, or adaptive representations~\cite{lotfollahi2019scgen,lotfollahi2023cpa,hetzel2022chemcpa,piran2024biolord,yang2026adapert,yao2026perturbedvae}.
These methods primarily use graph-derived relations within response computation or adapt cell- and condition-level representations without modifying gene-wise structural coordinates.
Spectral encoders such as SignNet provide global node coordinates but remain intervention agnostic~\cite{lim2023signnet}.
In contrast, \method{} derives multi-scale spectral coordinates from GO~\cite{ashburner2000geneontology} and control-derived coexpression, then uses perturbation-conditioned, gene-wise gates to select relevant scales and network sources.
The resulting geometry provides intervention-specific structural context without propagating target signals.

\paragraph{Geometry-conditioned residual flow.}
Unpaired population modeling has developed along two main lines.
Optimal-transport methods construct population couplings without cell-level correspondences: CellOT and CMonge learn conditional control-to-treatment maps, while CINEMA-OT reduces treatment-associated variation before matching~\cite{bunne2023cellot,driessen2026cmonge,dong2023cinemaot}.
Flow-based methods instead represent population change through a generative velocity field.
CellFlow combines OT coupling with conditional flow matching, whereas scDFM introduces graph-aware structure into distributional flow matching~\cite{klein2025cellflow,zhang2026scdfm}.
Despite their different parameterizations, these methods primarily address how a control population should be transported to a perturbed distribution, without explicitly separating stable gene structure from intervention-specific response dynamics.
\method{} restricts OT to endpoint construction and conditions a separate control-anchored, gene-wise residual field on $Z_p$.
Delta-correlation further aligns the predicted and observed condition-level expression shifts.

\section{Problem Formulation}

We define unseen-perturbation prediction as conditional population generation from unpaired single-cell observations.
Let $G$ be the number of modeled genes,
$x^{c}\sim\mathbb{P}_{c}$ a control cell, and
$y^{p}\sim\mathbb{P}_{p}$ a cell observed after perturbation $p$, all in $\mathbb{R}^{G}$.
Training data consist of finite control samples
$\mathcal{X}_{c}=\{x_i^{c}\}_{i=1}^{N_c}$ and perturbed samples
$\mathcal{Y}_{p}=\{y_j^{p}\}_{j=1}^{N_p}$ for
$p\in\mathcal{P}_{\mathrm{train}}$; destructive sequencing provides no cell-level correspondence between them.
Each condition is described by a nonempty gene-aligned target mask
$s_p\in\{0,1\}^{G}$, with $\mathcal{T}_p=\{g:s_{p,g}=1\}$.
Let $\mathcal{R}=\{\mathrm{GO},\mathrm{CE}\}$ and
$\mathcal{G}=\{W^{(r)}\}_{r\in\mathcal{R}}$ denote the fixed GO and control-coexpression graph priors.
Training and test conditions satisfy
$\mathcal{P}_{\mathrm{train}}\cap\mathcal{P}_{\mathrm{test}}=\varnothing$.
For an unseen $p^\star\in\mathcal{P}_{\mathrm{test}}$, the model receives only
$\mathcal{X}_{c}$, $s_{p^\star}$, and $\mathcal{G}$, with no sample from
$\mathbb{P}_{p^\star}$ available for fitting.

We seek a conditional generator
\begin{equation}
\begin{aligned}
\hat{y}^{p}
&\sim q_\theta(\,\cdot\mid x^{c},s_p,\mathcal{G}\,),\\
\widehat{\mathbb{P}}_{\theta,p}
&=\int q_\theta(\,\cdot\mid x^{c},s_p,\mathcal{G}\,)
\,d\mathbb{P}_{c}(x^{c}),
\end{aligned}
\label{eq:conditional_generator}
\end{equation}
where $\hat y^p$ is one generated cell and $\widehat{\mathbb{P}}_{\theta,p}$ is the predicted population after marginalizing over control cells.
For an unseen condition, the target is
$\widehat{\mathbb{P}}_{\theta,p^\star}\approx\mathbb{P}_{p^\star}$.
Only the marginalized population $\widehat{\mathbb{P}}_{\theta,p}$ is identifiable from unpaired observations; the conditional coupling represented by $q_\theta$ is a modeling choice and is not interpreted as an individual counterfactual map.
The model must therefore infer the response distribution of an unseen intervention from its target description and stable gene-structure priors, without assuming that those priors specify the response direction or magnitude.

\section{Method}

\method{} separates stable gene structure from intervention-specific response dynamics.
It constructs perturbation-conditioned gene geometry $Z_p$ by routing multi-scale coordinates from fixed GO and control-derived coexpression graphs.
Condition-wise OT supplies training endpoints, while a control-anchored gene-wise residual field uses $Z_p$ and perturbation context to predict responses; Delta correlation aligns condition-level shifts.
The default model uses graph geometry only as structural context, whereas \methodsp{} adds target-signal propagation as a matched control.
Figure~\ref{fig:method_overview} summarizes the workflow.

\subsection{Perturbation-Conditioned Gene Geometry}

\paragraph{Fixed biological coordinates.}
We construct two weighted graphs over the modeled genes: a GO graph from biological-process annotation overlap~\cite{ashburner2000geneontology} and a coexpression graph from absolute Pearson correlation across training control cells.
Let $\mathcal A_g$ be the GO term set of gene $g$ and $X^c$ the training-control expression matrix.
Their edge weights are
\begin{equation}
\begin{aligned}
w_{gh}^{\mathrm{GO}}
&=
\frac{|\mathcal A_g\cap\mathcal A_h|}
{|\mathcal A_g\cup\mathcal A_h|},\\
w_{gh}^{\mathrm{CE}}
&=
\left|\operatorname{Pearson}(X^c_{:,g},X^c_{:,h})\right|,
\end{aligned}
\label{eq:graph_weights}
\end{equation}
where $g$ and $h$ index genes.
Both graphs are sparsified, symmetrized, computed before training, and kept fixed; the exact sparsification rules are given in the supplementary material.
Rather than treating raw graph edges as response pathways, we use spectral coordinates to summarize each gene's higher-order structural position within each fixed graph.
For source $r\in\mathcal{R}=\{\mathrm{GO},\mathrm{CE}\}$, let $W^{(r)}$ be its adjacency and $D^{(r)}=\operatorname{diag}(W^{(r)}\mathbf{1})$.
We obtain degree-normalized spectral coordinates from
\begin{equation}
\begin{aligned}
L_{\mathrm{sym}}^{(r)}
&=
I-(D^{(r)})^{-1/2}W^{(r)}(D^{(r)})^{-1/2},\\
L_{\mathrm{sym}}^{(r)}U^{(r)}
&=U^{(r)}\Lambda^{(r)},\\
\Phi^{(r)}
&=(D^{(r)})^{-1/2}U^{(r)},
\end{aligned}
\label{eq:spectral_coordinates}
\end{equation}
where the columns of $U^{(r)}$ are Laplacian eigenvectors,
$\Lambda^{(r)}$ contains their eigenvalues, and $\Phi^{(r)}$ gives the degree-normalized gene coordinates; we write
$\lambda^{(r)}=\operatorname{diag}(\Lambda^{(r)})$ below.
After discarding the first trivial mode, we retain the next 32 eigenpairs.
We split them into low- and high-frequency blocks and apply separate SignNet-based Spectral Encoders (SPEs)~\cite{lim2023signnet}, obtaining fixed gene-wise features $b_{\mathrm{lo},g}^{(r)}$ and $b_{\mathrm{hi},g}^{(r)}$.
The low-frequency block represents coarse graph organization, whereas the high-frequency block retains finer structural variation; keeping both provides candidate scales for perturbation-conditioned routing.
Graph construction, isolated-node handling, and encoder dimensions are specified in the supplementary material.

\paragraph{Perturbation-conditioned gene geometry.}
A fixed coordinate system cannot express that interventions may depend on different structural scales and graph sources.
For the target set $\mathcal{T}_p$, let
$\bar b_g=[b_{\mathrm{lo},g}^{(r)};b_{\mathrm{hi},g}^{(r)}]_{r\in\mathcal{R}}$
and form the shared target-set embedding
$e_p^{\mathrm{geo}}=|\mathcal{T}_p|^{-1}\sum_{h\in\mathcal{T}_p}\phi_{\mathrm{geo}}(\bar b_h)$.
The Spectral Scale Router (SSR) mixes the two frequency blocks within each source, and the Graph Source Router (GSR) then fuses the sources:
\begin{equation}
\begin{aligned}
\alpha_{p,g}^{(r)}
&=\operatorname{sigmoid}\!\left(
f_r^{\mathrm{scale}}(
[b_{\mathrm{lo},g}^{(r)};b_{\mathrm{hi},g}^{(r)};e_p^{\mathrm{geo}}])
\right),\\
z_{p,g}^{(r)}
&=M_r\!\left(
\alpha_{p,g}^{(r)}b_{\mathrm{lo},g}^{(r)}
+(1-\alpha_{p,g}^{(r)})b_{\mathrm{hi},g}^{(r)}
\right),\\
\pi_{p,g}^{(r)}
&=\operatorname{softmax}_{r}\!\left(
f_r^{\mathrm{src}}([z_{p,g}^{(r)};e_p^{\mathrm{geo}}])
\right),\\
z_{p,g}
&=\sum_{r\in\mathcal{R}}\pi_{p,g}^{(r)}z_{p,g}^{(r)},
\end{aligned}
\label{eq:conditioned_gene_geometry}
\end{equation}
where $\phi_{\mathrm{geo}}$ is a shared target-gene encoder,
$f_r^{\mathrm{scale}}$ and $f_r^{\mathrm{src}}$ are learned scalar scorers, and $M_r$ maps each source to the common $d_z$-dimensional space.
In Equation~\eqref{eq:conditioned_gene_geometry}, $\alpha_{p,g}^{(r)}$ selects scale within each source, $\pi_{p,g}^{(r)}$ fuses sources, and stacking the outputs gives
$Z_p=[z_{p,1},\ldots,z_{p,G}]^\top\in\mathbb{R}^{G\times d_z}$.
Thus, interventions share the same structural basis while selecting different structural scales and graph sources for each gene.
The target mask conditions these selections and is not propagated along graph edges.

\subsection{Geometry-Conditioned Gene-Wise Residual Flow Matching}

Given the perturbation-conditioned geometry $Z_p$, the remaining task is to learn a conditional population transformation from unpaired control and perturbed observations.

\paragraph{Condition-wise cell coupling and control-anchored path.}
For condition $p$, equal-size control and perturbed minibatches are coupled by discrete optimal transport~\cite{dong2023cinemaot,klein2025cellflow,chi2026departures}:
\begin{equation}
\begin{aligned}
\rho_p^\star
&=
\underset{\rho\in\mathfrak{S}_B}{\arg\min}
\sum_{i=1}^{B}
\left\|x_i^c-y_{\rho(i)}^p\right\|_2^2,\\
y_i
&=
y_{\rho_p^\star(i)}^p,
\end{aligned}
\label{eq:ot_coupling}
\end{equation}
where $B$ is the condition-specific minibatch size,
$\mathfrak{S}_B$ is the set of index permutations, and $\rho_p^\star$ is the minimum-cost one-to-one assignment.
Compared with arbitrary pairing, this assignment reduces baseline-state discrepancies between endpoints so that residual targets are less dominated by cross-cell mismatch.
For each coupled pair, we perturb the measured control locally and construct a straight residual path:
\begin{equation}
\begin{alignedat}{2}
x_{i,0}
&=
x_i^c+\sigma\epsilon_i,
&\qquad
\epsilon_i
&\sim\mathcal{N}(0,I_G),\\
x_{i,t_i}
&=
(1-t_i)x_{i,0}+t_i y_i,
&\qquad
t_i
&\sim\mathcal{U}(0,1),\\
u_i
&=
y_i-x_{i,0}.&&
\end{alignedat}
\label{eq:residual_path}
\end{equation}
where $\sigma$ is the fixed control-anchor noise scale,
$x_{i,t_i}$ is the sampled path state, and $u_i$ is the constant target velocity along the straight path.
Perturbing the control anchor encourages a locally smooth velocity field, while independent noise draws provide sample diversity when generating populations.
The coupling and path are training constructions rather than lineage or individual counterfactual assignments.

\paragraph{Condition context.}
The response backbone processes genes with shared parameters; within this backbone, the pooled Condition Context is the only channel through which information is shared across genes.
It uses a separately parameterized perturbation embedding.
It summarizes routed target coordinates as
$e_p^{\mathrm{rsp}}=|\mathcal{T}_p|^{-1}\sum_{h\in\mathcal{T}_p}\phi_{\mathrm{rsp}}(z_{p,h})$.
A context encoder first constructs gene tokens and pools them into a shared condition representation:
\begin{equation}
\begin{aligned}
\eta_{i,p,g}
&=
[x_{i,g}^{c};z_{p,g};s_{p,g};e_p^{\mathrm{rsp}}],\\
\tau_{i,p,g}
&=
\operatorname{GeneTokenEncoder}(\eta_{i,p,g}),\\
\bar{\tau}_{i,p}
&=
\operatorname{AttnPool}
(\{\tau_{i,p,g}\}_{g=1}^{G}),\\
c_{i,p,t_i}
&=
[\bar{\tau}_{i,p};
\operatorname{TimeEmbed}(t_i);
e_p^{\mathrm{rsp}}],
\end{aligned}
\label{eq:condition_context}
\end{equation}
where $\phi_{\mathrm{rsp}}$ is a response-specific shared target-gene encoder,
GeneTokenEncoder maps each gene's inputs to a token, AttnPool forms a cell-level summary, and TimeEmbed applies a sinusoidal encoding followed by an MLP.
Learned-query attention pooling permits gene tokens to contribute unequally to the cell-level summary, while the time embedding identifies the current position along the control-to-perturbed path.
The resulting $c_{i,p,t_i}$ is the Condition Context in Figure~\ref{fig:method_overview}; it is shared across genes within cell $i$ but remains control-, perturbation-, and time-dependent.

\paragraph{Local projection.}
For each gene, we concatenate the interpolated expression, control anchor, perturbation-conditioned geometry, and target state:
\begin{equation}
\begin{aligned}
\xi_{i,p,t_i,g}
&=
[x_{i,t_i,g};x_{i,g}^{c};z_{p,g};s_{p,g}],\\
\ell_{i,p,t_i,g}
&=
P_{\mathrm{loc}}(\xi_{i,p,t_i,g})
=
\operatorname{SiLU}
(W_{\mathrm{loc}}\xi_{i,p,t_i,g}+b_{\mathrm{loc}}).
\end{aligned}
\label{eq:local_projection}
\end{equation}
The projection shares parameters across genes, while $z_{p,g}$ retains gene-specific structural context.

\paragraph{FiLM residual field and velocity head.}
The condition context first enters through an additive projection and then modulates residual blocks $k=1,\ldots,K$, with $K=3$.
To simplify notation, write $c=c_{i,p,t_i}$ and
$\ell_g=\ell_{i,p,t_i,g}$, and suppress the common indices $(i,p,t_i)$ inside the block recurrence.
\begin{equation}
\begin{aligned}
h_g^{(0)}
&=
\ell_g+P_{\mathrm{ctx}}(c),\\
\widetilde h_g^{(k)}
&=
h_g^{(k-1)}
+W_2^{(k)}
\operatorname{SiLU}\!\left(
W_1^{(k)}
\operatorname{LN}(h_g^{(k-1)})
\right),\\
[\gamma^{(k)};\beta^{(k)}]
&=
A^{(k)}c+a^{(k)},\\
h_g^{(k)}
&=
(1+\gamma^{(k)})\odot\widetilde h_g^{(k)}
+\beta^{(k)},\\
v_{\theta,i,g}
&=
w_{\mathrm{out}}^\top h_g^{(K)}
+b_{\mathrm{out}},
\end{aligned}
\label{eq:gene_velocity}
\end{equation}
where $P_{\mathrm{ctx}}$, $W_1^{(k)}$, $W_2^{(k)}$, and $A^{(k)}$ are learned maps;
$\operatorname{LN}$ denotes layer normalization, $\odot$ denotes element-wise multiplication, and
$\gamma^{(k)}$ and $\beta^{(k)}$ are condition-dependent FiLM scale and shift vectors.
Here $v_{\theta,i,g}$ abbreviates $v_{\theta,i,p,t_i,g}$ for the current $(p,t_i)$, and stacking over genes gives $v_i=(v_{\theta,i,g})_{g=1}^{G}$.
Bias terms in the residual maps are suppressed for readability.
FiLM treats $c$ as a shared control signal that adapts gene-local feature processing at each block through condition-dependent scale and shift parameters.
All residual-field and head parameters are shared over genes; genes interact only through the pooled Condition Context, not through graph propagation.

\paragraph{Flow-matching and Delta-correlation objectives.}
The flow-matching term fits the OT-coupled residual velocity:
\begin{equation}
\mathcal{L}_{\mathrm{FM}}
=
\frac{1}{BG}\sum_{i=1}^{B}\|v_i-u_i\|_2^2.
\label{eq:flow_matching_loss}
\end{equation}
This cell-level objective does not directly constrain the direction of the condition-level mean expression shift.
We therefore use the noise-free endpoint residual to impose a Delta-correlation objective across genes:
\begin{equation}
\begin{alignedat}{2}
\bar v_p
&=
\frac{1}{B}\sum_{i=1}^{B}v_i,
&\qquad
\bar d_p
&=
\frac{1}{B}\sum_{i=1}^{B}(y_i-x_i^c),\\
\mathcal{L}_{\Delta}
&=
1-\operatorname{corr}_{\kappa}(\bar v_p,\bar d_p).&&
\end{alignedat}
\label{eq:delta_correlation_loss}
\end{equation}
where $\operatorname{corr}_{\kappa}$ is a Pearson correlation computed across the $G$ genes with positive denominator floor $\kappa$.
The complete objective is
\begin{equation}
\mathcal{L}
=
\mathcal{L}_{\mathrm{FM}}
+\lambda_{\Delta}\mathcal{L}_{\Delta},
\label{eq:training_objective}
\end{equation}
where $\lambda_{\Delta}$ controls the strength of condition-level response alignment.
The stabilized correlation operator and layer dimensions are given in the supplementary material.

\paragraph{Explicit-propagation control.}
The default \method{} uses graph spectra only as structural coordinates and never spreads the perturbation mask along graph edges.
To isolate whether target-signal propagation adds value beyond this geometric conditioning, \methodsp{} introduces a matched learned low-rank spectral operator over the coexpression coordinates:
\begin{equation}
r_p^{(\ell)}
=
\Phi^{(\mathrm{CE})}
\operatorname{diag}\!\left(g_{\psi,\ell}(\lambda^{(\mathrm{CE})})\right)
(\Phi^{(\mathrm{CE})})^\top s_p.
\label{eq:explicit_propagation}
\end{equation}
where $\ell=1,\ldots,C$ indexes the learned propagation channels, $g_{\psi,\ell}$ is the filter for channel $\ell$, $\lambda^{(\mathrm{CE})}$ is the coexpression eigenvalue vector, and
$r_p^{(\ell)}\in\mathbb{R}^{G}$ is the resulting gene-wise propagated channel.
These channels are appended to the local input $\xi_{i,p,t_i,g}$ in Equation~\eqref{eq:local_projection}; all other components remain unchanged.
This is a learned low-rank projection and reconstruction rather than an assumed causal diffusion process.
Accordingly, the comparison evaluates this particular propagation operator under matched conditions rather than excluding all possible graph-propagation designs.

\paragraph{Inference.}
Inference requires neither OT assignment nor perturbed cells.
Given a control cell and an unseen target mask, the model constructs $Z_p$ and integrates
\begin{equation}
\begin{aligned}
\epsilon
&\sim\mathcal{N}(0,I_G),\\
x(0)
&=
x^c+\sigma\epsilon,\\
\frac{\mathrm{d}x(t)}{\mathrm{d}t}
&=
v_\theta
\left(
x(t),t,x^c,s_p,Z_p
\right),
\qquad
t\in[0,1].
\end{aligned}
\label{eq:inference_ode}
\end{equation}
Integrating Equation~\eqref{eq:inference_ode} from $t=0$ to $t=1$ yields the terminal state $x(1)=\hat y^p$, one generated perturbed cell.
Repeating this procedure over control cells and independent noise draws yields the predicted population $\widehat{\mathbb{P}}_{\theta,p}$.

\begin{table}[!t]
\centering
\caption{Norman additive results. Bold and underlined values denote the best and second-best results, respectively.}
\label{tab:main_results}
\begingroup
\setlength{\tabcolsep}{2pt}
\renewcommand{\arraystretch}{0.92}
\scriptsize
\begin{tabular*}{\columnwidth}{@{\extracolsep{\fill}}lrrr@{}}
\hline
Model & MSE $\downarrow$ & DE-Spearman $\rho$ $\uparrow$ & Pearson $\Delta$ $\uparrow$ \\
\hline
Control & 0.0184 & N.A. & N.A. \\
Additive & 0.0045 & 0.5564 & \textbf{0.9024} \\
Geneformer & 0.0041 & 0.3741 & 0.7732 \\
GEARS & 0.0139 & 0.5624 & 0.7421 \\
CPA & 0.0344 & 0.0713 & 0.3845 \\
STATE & 0.3006 & 0.5288 & $-0.0108$ \\
CellFlow & 0.0039 & 0.5503 & 0.8678 \\
scDFM & \underline{0.0032} & \underline{0.5705} & 0.8853 \\
\hline
\method{} & \textbf{0.0030} & \textbf{0.5992} & \underline{0.8979} \\
\hline
\end{tabular*}
\endgroup
\end{table}

\section{Experiments}

\subsection{Experimental Setup}

We evaluated generalization to unseen interventions and used matched controls to isolate the effects of perturbation-conditioned gene geometry and explicit graph propagation.

\paragraph{Datasets and evaluation protocols.}
Norman, our primary benchmark, contains K562 CRISPRa single- and two-gene interventions~\cite{norman2019exploring}.
Its five-fold strict-holdout protocol removes selected genes and every condition containing them from training, providing the main evidence for unseen genetic perturbations.
We also report the established additive and single/double holdout protocols for comparison with scDFM.
For cross-domain validation, we used ComboSciPlex, which contains A549 cells exposed to single agents and drug combinations~\cite{lotfollahi2023cpa}.
Its fixed split contains 24 training and seven held-out conditions; we focus on the five held-out drug--drug interventions and represent each combination by the union of its known targets.
Complete preprocessing, split, and graph statistics are provided in the supplementary material.

\paragraph{Baselines.}
Control reuses the control population and predicts no intervention shift.
Additive sums available component-wise condition-mean residuals.
Graph-free retains the same residual-flow backbone but removes spectral gene geometry, while \methodsp{} adds explicit target-signal propagation to \method{}.
The external tables additionally include scGPT~\cite{cui2024scgpt}, Geneformer~\cite{theodoris2023geneformer}, GEARS~\cite{roohani2023gears}, CPA~\cite{lotfollahi2023cpa}, STATE~\cite{adduri2025state}, CellFlow~\cite{klein2025cellflow}, and scDFM~\cite{zhang2026scdfm}.
For the Norman additive and single/double holdout comparisons, baseline values are transcribed from scDFM~\cite{zhang2026scdfm}, while the \method{} rows are obtained from our runs under the corresponding protocols.
Because we did not rerun these external methods, the benchmark tables provide cross-source context, whereas matched local controls support mechanism attribution.
\paragraph{Metrics and implementation.}
Our primary metrics were Pearson Delta, the gene-wise correlation between predicted and observed condition-mean residuals, and MSE, their numerical discrepancy.
Population fidelity was additionally measured by RBF maximum mean discrepancy (MMD) and the Pearson correlation between predicted and observed gene-wise variance vectors.
We averaged Norman results over data-split folds and ComboSciPlex results over conditions in its fixed test split unless stated otherwise.
All locally trained models used the same preprocessing, evaluation interface, and training budget.
Additional implementation and evaluation details are provided in the supplementary material.

\subsection{Quantitative Results on Norman}

\begin{figure*}[!t]
\centering
\includegraphics[width=0.88\textwidth]{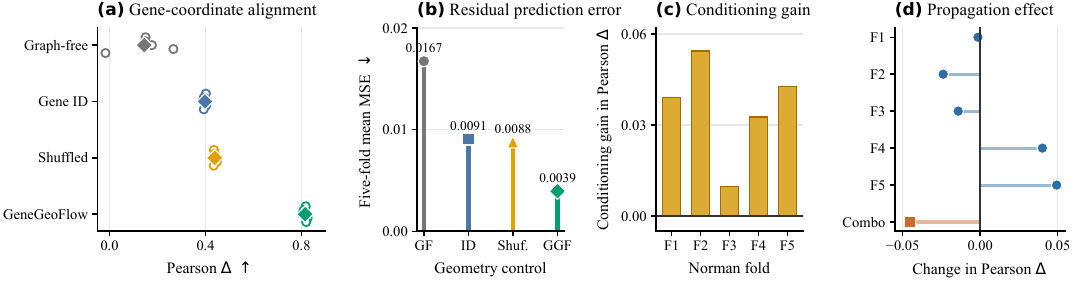}
\caption{Matched controls isolate the contributions of biological gene coordinates and perturbation conditioning from explicit graph propagation. Panels (a)--(b) compare Graph-free, learned gene IDs, shuffled coordinates, and aligned target-set GeneGeoFlow on five Norman strict-holdout folds; in panel (a), hollow markers denote folds and filled diamonds denote five-fold means, while panel (b) reports mean MSE. Panel (c) reports Conditioned minus Static per Norman fold. Panel (d) reports GeneGeoFlow-Prop minus GeneGeoFlow for Norman folds and the ComboSciPlex mean. Each panel is a within-family comparison: panels (a)--(b) use target-set pooling, panel (c) uses the matched router run family, and panel (d) uses the dense full-mask response-encoder run family.}
\label{fig:mechanism_ablation}
\end{figure*}

\begin{table}[!t]
\centering
\caption{Norman single- and double-perturbation holdout results. Bold and underlined values denote the best and second-best results within each setting, respectively.}
\label{tab:norman_holdout_external}
\begingroup
\setlength{\tabcolsep}{2pt}
\renewcommand{\arraystretch}{0.90}
\scriptsize
\begin{tabular*}{\columnwidth}{@{\extracolsep{\fill}}llrrr@{}}
\hline
Setting & Model & MSE $\downarrow$ & DE-Spearman $\rho$ $\uparrow$ & Pearson $\Delta$ $\uparrow$ \\
\hline
Single & Control & 0.0095 & N.A. & N.A. \\
& scGPT & 0.0080 & $-0.1139$ & 0.4503 \\
& GEARS & 0.0075 & 0.3569 & 0.6646 \\
& Geneformer & 0.0036 & 0.3669 & 0.6955 \\
& CPA & 0.0356 & 0.1168 & 0.2837 \\
& STATE & 0.3333 & 0.6116 & 0.0004 \\
& CellFlow & 0.0035 & 0.2860 & 0.7109 \\
& scDFM & \underline{0.0030} & \underline{0.6957} & \underline{0.7127} \\
& \method{} & \textbf{0.0029} & \textbf{0.7078} & \textbf{0.7421} \\
\hline
Double & Control & 0.0207 & N.A. & N.A. \\
& scGPT & 0.0153 & $-0.0665$ & 0.5693 \\
& GEARS & 0.0156 & 0.2543 & 0.7552 \\
& Geneformer & 0.0050 & 0.3468 & 0.7361 \\
& CPA & 0.0357 & 0.3652 & 0.4176 \\
& STATE & 0.3404 & 0.4071 & 0.0061 \\
& CellFlow & 0.0049 & 0.5074 & 0.8095 \\
& scDFM & \underline{0.0047} & \underline{0.5676} & \underline{0.8357} \\
& \method{} & \textbf{0.0044} & \textbf{0.5918} & \textbf{0.8464} \\
\hline
\end{tabular*}
\endgroup
\end{table}

Table~\ref{tab:norman_holdout_external} provides benchmark context on the Norman single- and double-perturbation holdouts.
\method{} outperformed scDFM in both settings, reaching Pearson Delta scores of $0.7421$ and $0.8464$, with MSE values of $0.0029$ and $0.0044$, on the single- and double-perturbation holdouts, respectively.

On the primary five-fold strict-holdout protocol, target-set GeneGeoFlow improved over Graph-free in every fold, increasing mean Pearson Delta from $0.1450$ to $0.8153$ and reducing MSE from $0.0167$ to $0.0039$.

\begin{figure*}[!t]
\centering
\includegraphics[width=0.90\textwidth]{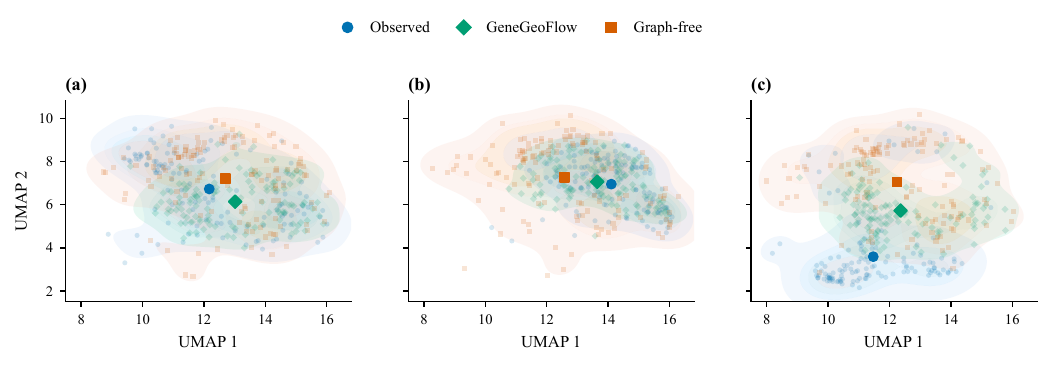}
\caption{Qualitative population-level comparison on three fold-1 Norman strict-holdout conditions: ZBTB10+PTPN12 (a), BAK1+control (b), and DUSP9+PRTG (c). PCA and UMAP are fitted to a fixed stratified sample of real cells and then used to project both predicted populations. Points and contours denote cells and density; enlarged markers denote centroids. Aggregate population fidelity is evaluated separately in Table~\ref{tab:population_fidelity}.}
\label{fig:population_umap}
\end{figure*}

The three conditions in Figure~\ref{fig:population_umap} span different levels of prediction difficulty, from partial recovery to close distributional agreement.
Geometry-conditioned predictions follow the observed density more closely in panels (b) and (c), whereas panel (a) shows partial recovery; Table~\ref{tab:population_fidelity} provides the quantitative population-level test.

Table~\ref{tab:main_results} reports results under the established Norman additive protocol.
\method{} achieved $0.8979$ Pearson Delta, compared with $0.9024$ for Additive and $0.8853$ for scDFM, while also reporting the lowest MSE and the highest DE-Spearman $\rho$.

\subsection{Quantitative Results on ComboSciPlex}

\begin{table}[!t]
\centering
\caption{ComboSciPlex results. Bold and underlined values denote the best and second-best results, respectively.}
\label{tab:combo_external_results}
\begingroup
\setlength{\tabcolsep}{2pt}
\renewcommand{\arraystretch}{0.92}
\scriptsize
\begin{tabular*}{\columnwidth}{@{\extracolsep{\fill}}lrrr@{}}
\hline
Model & MSE $\downarrow$ & DE-Spearman $\rho$ $\uparrow$ & Pearson $\Delta$ $\uparrow$ \\
\hline
Control & 0.0324 & N.A. & N.A. \\
scGPT & 0.0031 & $-0.1261$ & 0.8322 \\
CPA & 0.0029 & 0.7906 & 0.8150 \\
scDFM & \underline{0.0028} & \underline{0.8289} & \underline{0.8933} \\
\method{} & \textbf{0.0021} & \textbf{0.8305} & \textbf{0.9088} \\
\hline
\end{tabular*}
\endgroup
\end{table}

\begin{table}[!ht]
\centering
\caption{Population fidelity on held-out perturbations. Lower MMD and higher variance correlation are better; bold denotes the best result within each dataset.}
\label{tab:population_fidelity}
\begingroup
\renewcommand{\arraystretch}{0.90}
\scriptsize
\begin{tabular*}{\columnwidth}{@{\extracolsep{\fill}}llrr@{}}
\hline
Dataset & Variant & MMD $\downarrow$ & Var.\ Corr. $\uparrow$ \\
\hline
Norman & Graph-free & 0.1488 & 0.3114 \\
& \methodsp{} & 0.0965 & 0.6043 \\
& \method{} & \textbf{0.0821} & \textbf{0.6728} \\
ComboSciPlex & Graph-free & 0.1639 & 0.2862 \\
& \methodsp{} & 0.1011 & 0.5811 \\
& \method{} & \textbf{0.0894} & \textbf{0.6435} \\
\hline
\end{tabular*}
\endgroup
\end{table}

Table~\ref{tab:combo_external_results} places the ComboSciPlex result against reported baselines.
\method{} achieved the best results on all three reported metrics (MSE, DE-Spearman $\rho$, Pearson $\Delta$), including the strongest Pearson Delta ($0.9088$ versus $0.8933$ for scDFM).
In the matched local comparison, \method{} improved Pearson Delta from $0.5169$ to $0.9088$ over Graph-free and reduced MSE from $0.0733$ to $0.0021$; explicit propagation reached $0.8636$ and $0.0198$.
\method{} improved all five unseen combinations by $0.1900$--$0.4800$, including $0.2700$ to $0.7100$ on SRT2104+Alvespimycin.
\method{} also achieved the lowest MMD and the highest variance correlation on both datasets (Table~\ref{tab:population_fidelity}).

\subsection{Ablation Study}

\paragraph{Explicit-propagation control.}
Figure~\ref{fig:mechanism_ablation} separates structural coordinates from explicit target-signal propagation.
Because the propagation analysis uses a dense full-mask response encoder, all conclusions in this panel are drawn within the same run family; its no-propagation reference therefore differs from the final target-set model.
The evaluated low-rank operator produced mixed fold-wise effects on Norman, with a mean Pearson Delta change of $+0.0101$, while decreasing ComboSciPlex performance by $0.0452$.
It therefore provided no consistent benefit in this setting, without ruling out other propagation designs.

\paragraph{Perturbation-conditioned geometry.}
Static geometry preserves the biological coordinates and target mask but removes the perturbation input from both routers.
Conditioning improved every matched Norman fold, with a mean Pearson Delta gain of $0.0357$ (Figure~\ref{fig:mechanism_ablation}(c)), isolating intervention-dependent scale and source selection.

\paragraph{Gene coordinate alignment.}
With the response backbone fixed, Learned Gene ID, Shuffled Geometry, aligned geometry, and Graph-free reached mean Pearson Delta values of $0.3981$, $0.4387$, $0.8153$, and $0.1450$, respectively (Figure~\ref{fig:mechanism_ablation}(a)).
Their mean MSE values were $0.0091$, $0.0088$, $0.0039$, and $0.0167$ (Figure~\ref{fig:mechanism_ablation}(b)), indicating that the gain cannot be explained by gene identity, input dimensionality, or arbitrary coordinate assignment alone.

\FloatBarrier
\section{Conclusion}

We introduced \method{} to separate stable gene structure from intervention-specific response dynamics.
It constructs perturbation-conditioned gene geometry from GO and control-derived coexpression networks to condition a control-anchored gene-wise residual flow; condition-wise OT and Delta correlation address unpaired populations and condition-level alignment.
Matched comparisons on Norman and ComboSciPlex attributed the main gains to biologically aligned coordinates and perturbation-conditioned routing, whereas the evaluated low-rank propagation operator yielded no consistent benefit.
These results support using biological networks as structural conditions for data-driven response fields.

\clearpage
\bibliography{aaai2027}

\clearpage
\appendix
\setcounter{secnumdepth}{1}
\section*{Supplementary Material}
\setcounter{equation}{0}
\setcounter{table}{0}
\renewcommand{\theequation}{S\arabic{equation}}
\renewcommand{\thetable}{S\arabic{table}}

\section{Additional Method Details}

This supplement records the construction and implementation details omitted from the condensed method in the main paper.
It uses the same notation as the main paper and introduces no additional model component.

\subsection{Graph Construction and Spectral Coordinates}

\paragraph{Fixed graph priors.}
Let the training control population be
$X^c=[x_1^c,\ldots,x_{N_c}^c]^\top\in\mathbb{R}^{N_c\times G}$,
and let $\mathcal{A}_g$ be the set of GO biological-process terms assigned to gene $g$.
The two graph weights are
\begin{equation}
\begin{aligned}
w_{gh}^{\mathrm{GO}}
&=
\frac{|\mathcal{A}_g\cap\mathcal{A}_h|}
{|\mathcal{A}_g\cup\mathcal{A}_h|},\\
w_{gh}^{\mathrm{CE}}
&=
\left|
\operatorname{Pearson}(X^c_{:,g},X^c_{:,h})
\right|.
\end{aligned}
\label{eq:supp_graph_weights}
\end{equation}
Equation~\eqref{eq:supp_graph_weights} is evaluated using training controls only.
For GO, we retain the 20 highest-weight neighbors of each gene.
For coexpression, we retain edges with absolute correlation at least $0.3$ and augment them with the top 20 neighbors of each gene.
The directed top-$k$ matrices are symmetrized by arithmetic averaging, $W\leftarrow(W+W^\top)/2$, and fixed before model training.
Perturbed cells are not used to construct either graph.
GO annotations are read from the GAF 2.2 biological-process namespace using gene symbols; annotations with the \texttt{NOT} qualifier are excluded.
The archived annotation file was generated on March 28, 2026 from the March 1, 2026 GO release.

\paragraph{Spectral data.}
For $r\in\{\mathrm{GO},\mathrm{CE}\}$, define
$D^{(r)}=\operatorname{diag}(W^{(r)}\mathbf{1})$.
The complete fixed spectral construction is
\begin{equation}
\begin{aligned}
L_{\mathrm{sym}}^{(r)}
&=
I-(D^{(r)})^{-1/2}W^{(r)}(D^{(r)})^{-1/2},\\
L_{\mathrm{sym}}^{(r)}U^{(r)}
&=
U^{(r)}\Lambda^{(r)},
\qquad
(U^{(r)})^\top U^{(r)}=I,\\
\Phi^{(r)}
&=
(D^{(r)})^{-1/2}U^{(r)},\\
\mathcal{S}
&=
\{(\Phi^{(r)},\lambda^{(r)})\}_{r\in\{\mathrm{GO},\mathrm{CE}\}} .
\end{aligned}
\label{eq:supp_spectral_data}
\end{equation}
Equation~\eqref{eq:supp_spectral_data} is computed once for each graph source and cached.
For isolated genes, the corresponding entry of $(D^{(r)})^{-1/2}$ is set to zero.
We discard the first trivial mode and retain the next 32 eigenpairs, ordered by increasing eigenvalue.
The first 16 retained modes form the low-frequency block and the remaining 16 form the high-frequency block.
Separate SignNet encoders resolve eigenvector sign ambiguity in the two blocks and produce
$b_{\mathrm{lo},g}^{(r)}$ and $b_{\mathrm{hi},g}^{(r)}$.
The eigendecompositions are performed once and cached.

\subsection{Perturbation-Conditioned Routing}

Let the nonempty target set be $\mathcal{T}_p=\{g:s_{p,g}=1\}$, and let
$\bar b_g=[b_{\mathrm{lo},g}^{(r)};b_{\mathrm{hi},g}^{(r)}]_{r\in\{\mathrm{GO},\mathrm{CE}\}}$
collect the fixed multi-scale coordinates of gene $g$.
The geometry embedding and source-specific routing variables are
\begin{equation}
\begin{aligned}
e_p^{\mathrm{geo}}
&=
\frac{1}{|\mathcal{T}_p|}
\sum_{h\in\mathcal{T}_p}
\phi_{\mathrm{geo}}(\bar b_h),\\
q_{p,g}^{(r)}
&=[b_{\mathrm{lo},g}^{(r)};
b_{\mathrm{hi},g}^{(r)};
e_p^{\mathrm{geo}}],\\
\alpha_{p,g}^{(r)}
&=\operatorname{sigmoid}
\left(f_r^{\mathrm{scale}}(q_{p,g}^{(r)})\right),\\
\widetilde z_{p,g}^{(r)}
&=
\alpha_{p,g}^{(r)}b_{\mathrm{lo},g}^{(r)}
+(1-\alpha_{p,g}^{(r)})b_{\mathrm{hi},g}^{(r)},\\
z_{p,g}^{(r)}
&=M_r\widetilde z_{p,g}^{(r)}.
\end{aligned}
\label{eq:supp_scale_router}
\end{equation}
After aligning the source-specific spaces, the Graph Source Router computes
\begin{equation}
\begin{aligned}
a_{p,g}^{(r)}
&=
f_r^{\mathrm{src}}([z_{p,g}^{(r)};e_p^{\mathrm{geo}}]),\\
\pi_{p,g}^{(r)}
&=
\frac{\exp(a_{p,g}^{(r)})}
{\sum_{r'}\exp(a_{p,g}^{(r')})},\\
z_{p,g}
&=
\sum_r\pi_{p,g}^{(r)}z_{p,g}^{(r)},\\
Z_p
&=
[z_{p,1},\ldots,z_{p,G}]^\top.
\end{aligned}
\label{eq:supp_source_router}
\end{equation}
Equations~\eqref{eq:supp_scale_router} and~\eqref{eq:supp_source_router} define the two-stage routing operation.
The target set affects both gates through a shared coordinate encoder and mean pooling in $e_p^{\mathrm{geo}}$.
It does not alter the graph eigensystems or diffuse along graph edges.

\subsection{Residual Velocity Backbone}

\paragraph{Shared condition context.}
The geometry and response target-set encoders use separate parameters.
For sample $i$, the response branch constructs
\begin{equation}
\begin{aligned}
e_p^{\mathrm{rsp}}
&=
\frac{1}{|\mathcal{T}_p|}
\sum_{h\in\mathcal{T}_p}
\phi_{\mathrm{rsp}}(z_{p,h}),\\
\eta_{i,p,g}
&=
[x_{i,g}^c;z_{p,g};s_{p,g};e_p^{\mathrm{rsp}}],\\
\tau_{i,p,g}
&=
\operatorname{GeneTokenEncoder}(\eta_{i,p,g}),\\
\bar{\tau}_{i,p}
&=
\operatorname{AttnPool}(\{\tau_{i,p,g}\}_{g=1}^{G}),\\
c_{i,p,t_i}
&=
[\bar{\tau}_{i,p};
\operatorname{TimeEmbed}(t_i);
e_p^{\mathrm{rsp}}].
\end{aligned}
\label{eq:supp_response_context}
\end{equation}
Equation~\eqref{eq:supp_response_context} defines the shared condition context.
The GeneTokenEncoder contains two linear--normalization--GELU layers.
Attention pooling supplies a shared cell-level context, while gene identity remains explicit in the local branch.
Because $\phi_{\mathrm{geo}}$ and $\phi_{\mathrm{rsp}}$ are shared across genes, these embeddings remain defined for target genes that are absent as intervention targets during training.

\paragraph{Local projection and FiLM residual field.}
For every gene,
\begin{equation}
\begin{aligned}
\ell_{i,p,t_i,g}
&=
P_{\mathrm{loc}}
([x_{i,t_i,g};x_{i,g}^c;z_{p,g};s_{p,g}]),\\
v_{\theta,i,g}
&=
\operatorname{VelocityHead}
\left(
\operatorname{FiLMRes}
(\ell_{i,p,t_i,g};c_{i,p,t_i})
\right),\\
v_i
&=
[v_{\theta,i,1},\ldots,v_{\theta,i,G}]^\top.
\end{aligned}
\label{eq:supp_gene_velocity}
\end{equation}
Here $v_{\theta,i,g}$ abbreviates $v_{\theta,i,p,t_i,g}$ for the current condition and sampled time.
The implemented backbone uses three residual blocks of width 256 and outputs one scalar velocity per gene.
The default model contains no graph message-passing layer.

\subsection{Training Objective and Numerical Stabilization}

Within each condition, we form the squared-Euclidean cost matrix between sampled control and perturbed cells and solve an exact balanced linear assignment with the Hungarian algorithm.
No entropic regularization or Sinkhorn approximation is used.
For the resulting OT-coupled minibatch, the flow-matching term is
\begin{equation}
\mathcal{L}_{\mathrm{FM}}
=
\frac{1}{BG}
\sum_{i=1}^{B}\|v_i-u_i\|_2^2,
\qquad
u_i=y_i-(x_i^c+\sigma\epsilon_i).
\label{eq:supp_fm_loss}
\end{equation}
Equation~\eqref{eq:supp_fm_loss} uses the noise-perturbed source only in the flow target.
The Delta target excludes source noise:
\begin{equation}
\bar v_p
=
\frac{1}{B}\sum_{i=1}^{B}v_i,
\qquad
\bar d_p
=
\frac{1}{B}\sum_{i=1}^{B}(y_i-x_i^c).
\label{eq:supp_delta_statistics}
\end{equation}
Equation~\eqref{eq:supp_delta_statistics} instead forms the condition-level target from noise-free residuals.
For $a,b\in\mathbb{R}^{G}$, define
\begin{equation}
\begin{aligned}
\mathsf{C}(a)
&=
a-\frac{1}{G}(\mathbf{1}^{\top}a)\mathbf{1},\\
d_{\kappa}(a,b)
&=
\max\left(
\|\mathsf{C}(a)\|_2\|\mathsf{C}(b)\|_2,
\kappa
\right),\\
\operatorname{corr}_{\kappa}(a,b)
&=
\frac{\langle\mathsf{C}(a),\mathsf{C}(b)\rangle}
{d_{\kappa}(a,b)},\\
\mathcal{L}_{\Delta}
&=
1-\operatorname{corr}_{\kappa}(\bar v_p,\bar d_p),\\
\mathcal{L}
&=
\mathcal{L}_{\mathrm{FM}}
+\lambda_{\Delta}\mathcal{L}_{\Delta}.
\end{aligned}
\label{eq:supp_delta_loss}
\end{equation}
Equation~\eqref{eq:supp_delta_loss} combines stabilized Delta correlation with flow matching.
We use $\kappa=10^{-8}$; the denominator floor affects only numerically degenerate batches.

\subsection{Explicit-Propagation Control}

The default model uses the graph spectra only to form $Z_p$.
The matched \methodsp{} control additionally constructs $C$ learned channels over the coexpression coordinates:
\begin{equation}
\begin{aligned}
r_p^{(\ell)}
&=
\Phi^{(\mathrm{CE})}
\operatorname{diag}
\left(
g_{\psi,\ell}(\lambda^{(\mathrm{CE})})
\right)\\
&\quad\cdot
(\Phi^{(\mathrm{CE})})^\top s_p,
\qquad \ell=1,\ldots,C.
\end{aligned}
\label{eq:supp_explicit_propagation}
\end{equation}
The channels in Equation~\eqref{eq:supp_explicit_propagation} are appended to the local features in
Equation~\eqref{eq:supp_gene_velocity}; the geometry routers, context encoder, velocity backbone, training coupling, and objectives remain unchanged.
Because $\Phi^{(\mathrm{CE})}$ contains degree-normalized random-walk coordinates rather than an orthonormal Euclidean basis, this operator is a learned low-rank projection and reconstruction, not an exact causal diffusion process.

\subsection{Implementation and Evaluation Settings}

\paragraph{Datasets and preprocessing.}
We remove empty cells and genes, total-count normalize non-log-transformed matrices to $10^4$ counts per cell, and apply \texttt{log1p}.
The modeled space contains the 5,000 highly variable genes selected with the Scanpy \texttt{seurat} procedure together with every perturbation target present in the dataset.
Genetic interventions are represented by multi-hot target masks, while a drug combination uses the union of the mapped targets of its components.
Table~\ref{tab:supp_data_graph_stats} reports the resulting modeled spaces and fixed graph sizes.

\begin{table}[!ht]
\centering
\caption{Dataset and fixed-graph statistics used in the reported runs. Edge counts refer to undirected edges after symmetrization.}
\label{tab:supp_data_graph_stats}
\scriptsize
\begin{tabular*}{\columnwidth}{@{\extracolsep{\fill}}lrrrrr@{}}
\toprule
Dataset & Genes & Controls & Cond. & GO edges & CE edges \\
\midrule
Norman & 5,029 & 7,275 & 226 & 63,755 & 66,406 \\
ComboSciPlex & 5,028 & 1,451 & 31 & 30,365 & 50,818 \\
\bottomrule
\end{tabular*}
\end{table}

\paragraph{Evaluation splits.}
Table~\ref{tab:supp_split_counts} gives the condition counts used for optimization and evaluation.
For the Norman strict holdout, each fold withholds 12 genes and assigns every condition containing at least one withheld gene to the test set; fold construction uses seed $42+\text{fold index}$.
For ComboSciPlex, the released fixed split contains 24 training and seven test conditions, and one training condition is reserved for validation.
The five reported drug--drug conditions are Panobinostat+Crizotinib, Panobinostat+Curcumin, Panobinostat+SRT1720, Panobinostat+Sorafenib, and SRT2104+Alvespimycin.
The other two fixed-split test conditions are control+Alvespimycin and control+Dacinostat.

\begin{table}[!ht]
\centering
\caption{Numbers of training, validation, and test conditions after validation construction. The Norman additive counts apply to each of its five folds.}
\label{tab:supp_split_counts}
\scriptsize
\begin{tabular*}{\columnwidth}{@{\extracolsep{\fill}}llrrr@{}}
\toprule
Protocol & Fold & Train & Val. & Test \\
\midrule
Norman additive & 1--5 & 180 & 9 & 37 \\
Norman strict & 1 & 172 & 9 & 45 \\
 & 2 & 180 & 10 & 36 \\
 & 3 & 185 & 10 & 31 \\
 & 4 & 179 & 9 & 38 \\
 & 5 & 166 & 9 & 51 \\
ComboSciPlex fixed & -- & 23 & 1 & 7 \\
\bottomrule
\end{tabular*}
\end{table}

\begin{table}[!ht]
\centering
\caption{Architecture settings used in the reported runs.}
\label{tab:supp_architecture}
\scriptsize
\begin{tabular*}{\columnwidth}{@{\extracolsep{\fill}}ll}
\toprule
Component & Setting \\
\midrule
GO graph & Biological process; top 20 \\
Coexpression graph & Control only; top 20; threshold 0.3 \\
Retained spectra & 32 per graph after first mode \\
Frequency split & 16 low / 16 high \\
Target-set embeddings & 32 / 32; mean pool; separate parameters \\
Geometry width $d_z$ & 64 \\
Gene-token width & 128 \\
Velocity backbone & Width 256; 3 residual blocks \\
Propagation channels & 0 default; 8 in \methodsp{} \\
\bottomrule
\end{tabular*}
\end{table}

\begin{table}[!ht]
\centering
\caption{Optimization and sampling settings used in the reported runs.}
\label{tab:supp_training}
\scriptsize
\begin{tabular*}{\columnwidth}{@{\extracolsep{\fill}}ll}
\toprule
Component & Setting \\
\midrule
Control noise $\sigma$ & 0.2 \\
Optimizer & AdamW; weight decay $10^{-5}$ \\
Learning rate & $3\times10^{-4}$ \\
Schedule & 2,000-step warmup; cosine decay \\
Maximum steps & 200,000 \\
Per-condition sampling & 256 cells \\
Maximum batch size & 512, grouped by condition \\
Gradient clipping & 1.0 \\
Delta weight $\lambda_{\Delta}$ & 0.03 \\
EMA decay & 0.999 \\
Validation interval & 5,000 steps \\
Model selection & Best validation Pearson Delta \\
Inference solver & Euler; 30 steps \\
Benchmark export & 128 controls; one draw per control \\
Population evaluation & 128 controls; 10 draws per control \\
\bottomrule
\end{tabular*}
\end{table}

\paragraph{Hyperparameter exploration.}
On Norman fold 1 with seed 42, we evaluated
$\lambda_{\Delta}\in\{0.01,0.03,0.05\}$,
$\sigma\in\{0.15,0.20,0.25\}$,
propagation channel counts in $\{4,8,16\}$,
propagation scales in $\{0,0.25,0.5,1.0\}$, and
inference control counts in $\{128,256\}$.
We selected the final settings by validation Pearson Delta and then fixed them for the reported runs.

\paragraph{Reproducibility.}
We use seed 42 for Python, NumPy, and PyTorch, including condition splitting, control-cell sampling, and evaluation.
The validation set is drawn from the training conditions only, and model selection uses the exponential-moving-average parameters at the checkpoint with the highest validation Pearson Delta.
Because deterministic CUDA algorithms are not forced, exact bitwise equality can depend on the hardware and software stack.

\paragraph{Computing environment.}
The primary experiments were executed on an Ubuntu 20.04.6 server with eight NVIDIA H20 GPUs (96\,GB memory per GPU), two Intel Xeon Platinum 8480+ processors (224 logical CPU threads in total), and 1.8\,TiB system memory.
Each training process was bound to one H20 GPU, while independent folds and ablations were distributed across the eight devices.
The software environment used Python 3.10.20, PyTorch 2.7.1 with CUDA 11.8 and cuDNN 9.1, NumPy 1.26.4, SciPy 1.15.3, AnnData 0.11.4, Scanpy 1.11.5, pandas 2.3.3, and scikit-learn 1.7.2; the server used NVIDIA driver 570.124.06.

\paragraph{Evaluation scope and aggregation.}
The benchmark tables use the same 1,000-gene evaluation interface as their corresponding external protocols.
Norman values are unweighted means over data-split folds, whereas ComboSciPlex values are fixed-split point estimates unless stated otherwise.
Within an evaluation, condition-level metrics are first computed separately and then averaged with equal weight across conditions.
Population MMD uses the biased estimator, including diagonal terms, with the summed RBF kernel
$k(x,y)=\sum_{h\in\mathcal{H}}\exp(-\|x-y\|_2^2/(2h^2))$ and
$\mathcal{H}=\{0.1,0.5,1.0,5.0\}$.
Variance correlation compares predicted and observed gene-wise variances.

\paragraph{Inference.}
OT assignments and perturbed cells are used only during training.
At inference, each sampled control is initialized as $x^{(0)}=x^c+\sigma\epsilon$ and advanced through the learned velocity field for 30 Euler steps; independent controls and noise draws form the generated population.

\begin{table*}[!t]
\centering
\caption{Comparison of reconstruction, differential-expression, and response-shift metrics on the Norman additive split. Bold and underlined values denote the best and second-best results, respectively.}
\label{tab:supp_norman_additive_full}
\begingroup
\setlength{\tabcolsep}{2pt}
\renewcommand{\arraystretch}{0.92}
\scriptsize
\begin{tabular*}{\textwidth}{@{\extracolsep{\fill}}lrrrrrrrr@{}}
\toprule
Model & L2 $\downarrow$ & MSE $\downarrow$ & MAE $\downarrow$ &
DE-Spear. $\rho$ $\uparrow$ & Pearson $\Delta$ $\uparrow$ & DS $\uparrow$ &
Pearson $\widehat{\Delta}$ $\uparrow$ & Pearson $\widehat{\Delta}_{20}$ $\uparrow$ \\
\midrule
Control & 3.9937 & 0.01839 & 0.03953 & N.A. & N.A. & 0.5135 & $-0.1695$ & $-0.1297$ \\
Additive & 1.9395 & 0.00448 & 0.02276 & 0.5564 & \textbf{0.9024} & 0.9686 & \underline{0.8584} & \underline{0.9244} \\
scGPT & 3.4112 & 0.01349 & 0.03796 & $1.07{\times}10^{-5}$ & 0.5304 & 0.5404 & 0.2165 & 0.2414 \\
Geneformer & 1.9132 & 0.00410 & 0.02360 & 0.3741 & 0.7732 & 0.8241 & $-0.0078$ & 0.2239 \\
GEARS & 3.5531 & 0.01387 & 0.06624 & 0.5624 & 0.7421 & 0.8601 & $-0.0089$ & 0.2032 \\
CPA & 5.7629 & 0.03435 & 0.07894 & 0.0713 & 0.3845 & 0.6021 & $-0.0039$ & 0.2254 \\
STATE & 17.3330 & 0.30059 & 0.24705 & 0.5288 & $-0.0108$ & 0.5135 & $-0.0069$ & 0.2515 \\
CellFlow & 1.7064 & 0.00392 & 0.02207 & 0.5503 & 0.8678 & 0.9321 & 0.8395 & 0.8988 \\
scDFM & \underline{1.7043} & \underline{0.00315} & \underline{0.02155} & \underline{0.5705} & 0.8853 & \underline{0.9737} & 0.8468 & \textbf{0.9260} \\
\midrule
\method{} & \textbf{1.6923} & \textbf{0.00303} & \textbf{0.02070} & \textbf{0.5992} & \underline{0.8979} & \textbf{0.9775} & \textbf{0.8601} & \textbf{0.9260} \\
\bottomrule
\end{tabular*}
\endgroup
\end{table*}

\begin{table*}[!t]
\centering
\caption{Comparison of reconstruction, differential-expression, and response-shift metrics on the Norman single- and double-perturbation holdout splits. Bold and underlined values denote the best and second-best results within each setting, respectively.}
\label{tab:supp_norman_holdout_full}
\begingroup
\setlength{\tabcolsep}{1.6pt}
\renewcommand{\arraystretch}{0.90}
\scriptsize
\begin{tabular*}{\textwidth}{@{\extracolsep{\fill}}llrrrrrrrr@{}}
\toprule
Setting & Model & L2 $\downarrow$ & MSE $\downarrow$ & MAE $\downarrow$ &
DE-Spear. $\rho$ $\uparrow$ & Pearson $\Delta$ $\uparrow$ & DS $\uparrow$ &
Pearson $\widehat{\Delta}$ $\uparrow$ & Pearson $\widehat{\Delta}_{20}$ $\uparrow$ \\
\midrule
Single & Control & 2.6834 & 0.0095 & 0.0263 & N.A. & N.A. & 0.5217 & 0.1618 & 0.1982 \\
& scGPT & 2.5007 & 0.0080 & 0.0259 & $-0.1139$ & 0.4503 & 0.5680 & 0.0747 & 0.0798 \\
& GEARS & 2.5641 & 0.0075 & 0.0466 & 0.3569 & 0.6646 & 0.8271 & 0.6356 & 0.7914 \\
& Geneformer & 1.6962 & 0.0036 & 0.0191 & 0.3669 & 0.6955 & 0.8070 & 0.5620 & 0.6513 \\
& CPA & 5.8060 & 0.0356 & 0.0853 & 0.1168 & 0.2837 & 0.5796 & $-0.0028$ & 0.0802 \\
& STATE & 18.2543 & 0.3333 & 0.2693 & 0.6116 & 0.0004 & 0.5236 & 0.0154 & 0.2386 \\
& CellFlow & 1.6758 & 0.0035 & 0.0191 & 0.2860 & 0.7109 & 0.8072 & 0.6138 & 0.6753 \\
& scDFM & \underline{1.6186} & \underline{0.0030} & \underline{0.0190} & \underline{0.6957} & \underline{0.7127} & \underline{0.8914} & \underline{0.6659} & \underline{0.8116} \\
& \method{} & \textbf{1.5487} & \textbf{0.00291} & \textbf{0.01812} & \textbf{0.7078} & \textbf{0.7421} & \textbf{0.8956} & \textbf{0.6914} & \textbf{0.8197} \\
\midrule
Double & Control & 4.1882 & 0.0207 & 0.0423 & N.A. & N.A. & 0.5322 & $-0.1303$ & $-0.0265$ \\
& scGPT & 3.5171 & 0.0153 & 0.0362 & $-0.0665$ & 0.5693 & 0.5578 & 0.2814 & 0.2652 \\
& GEARS & 3.7458 & 0.0156 & 0.0708 & 0.2543 & 0.7552 & 0.8766 & 0.6407 & 0.8413 \\
& Geneformer & 2.0819 & 0.0050 & 0.0237 & 0.3468 & 0.7361 & 0.8067 & 0.6245 & 0.7261 \\
& CPA & 5.7891 & 0.0357 & 0.0796 & 0.3652 & 0.4176 & 0.6311 & 0.2432 & 0.2870 \\
& STATE & 18.4458 & 0.3404 & 0.2733 & 0.4071 & 0.0061 & 0.5289 & $-0.0023$ & 0.2580 \\
& CellFlow & 2.1042 & 0.0049 & 0.0236 & 0.5074 & 0.8095 & 0.8622 & 0.6780 & 0.7155 \\
& scDFM & \underline{2.0309} & \underline{0.0047} & \underline{0.0235} & \underline{0.5676} & \underline{0.8357} & \underline{0.9189} & \underline{0.7769} & \underline{0.8688} \\
& \method{} & \textbf{1.9732} & \textbf{0.00437} & \textbf{0.02306} & \textbf{0.5918} & \textbf{0.8464} & \textbf{0.9321} & \textbf{0.7867} & \textbf{0.8875} \\
\bottomrule
\end{tabular*}
\endgroup
\end{table*}

\begin{table*}[!t]
\centering
\caption{Comparison of reconstruction, differential-expression, and response-shift metrics on the five held-out drug--drug conditions in the fixed ComboSciPlex split. Bold and underlined values denote the best and second-best available results, respectively; dashes indicate unavailable metrics.}
\label{tab:supp_combo_full}
\begingroup
\setlength{\tabcolsep}{3pt}
\renewcommand{\arraystretch}{0.92}
\scriptsize
\begin{tabular*}{\textwidth}{@{\extracolsep{\fill}}lrrrrrr@{}}
\toprule
Model & L2 $\downarrow$ & MSE $\downarrow$ & MAE $\downarrow$ &
DE-Spear. $\rho$ $\uparrow$ & Pearson $\Delta$ $\uparrow$ & DS $\uparrow$ \\
\midrule
Control & 5.3716 & 0.0324 & 0.0698 & N.A. & N.A. & 0.5714 \\
scGPT & 1.6934 & 0.0031 & 0.0251 & $-0.1261$ & 0.8322 & 0.8571 \\
CPA & 1.6592 & 0.0029 & 0.0240 & 0.7906 & 0.8150 & \textbf{0.8980} \\
scDFM & \underline{1.6567} & \underline{0.0028} & \underline{0.0220} & \underline{0.8289} & \underline{0.8933} & \underline{0.8776} \\
\method{} & \textbf{1.6491} & \textbf{0.0021} & \textbf{0.0213} & \textbf{0.8305} & \textbf{0.9088} & 0.8772 \\
\bottomrule
\end{tabular*}
\endgroup
\end{table*}

\FloatBarrier

\noindent
Baseline rows follow the same external benchmark sources as the condensed tables in the main paper, while the \method{} rows report our runs under the corresponding protocols.

\end{document}